\documentclass[conference]{IEEEtran}
\IEEEoverridecommandlockouts
\usepackage{comment}

\usepackage{cite}
\usepackage{amsmath,amssymb,amsfonts}
\usepackage{algorithmic}
\usepackage{graphicx}
\usepackage{textcomp}
\usepackage{xcolor}
\def\BibTeX{{\rm B\kern-.05em{\sc i\kern-.025em b}\kern-.08em
    T\kern-.1667em\lower.7ex\hbox{E}\kern-.125emX}}

\usepackage{booktabs}
\usepackage{multirow}
\usepackage[caption=false,font=footnotesize]{subfig}

\begin{document}

\title{Towards Autonomous Commissioning of Industrial Drives via Multi-Objective Bayesian Optimization\\
\thanks{This work was partially supported by CMZ Sistemi Elettronici S.r.l.}
}

\author{\IEEEauthorblockN{1\textsuperscript{st} David Petrovic}
\IEEEauthorblockA{\textit{Department of Information Engineering} \\
\textit{University of Padua}\\
Padua, Italy \\
david.petrovic@phd.unipd.it}
\and
\IEEEauthorblockN{2\textsuperscript{nd} Gian Antonio Susto}
\IEEEauthorblockA{\textit{Department of Information Engineering} \\
\textit{University of Padua}\\
Padua, Italy \\
gianantonio.susto@unipd.it}
\and
\IEEEauthorblockN{3\textsuperscript{rd} Angelo Cenedese}
\IEEEauthorblockA{\textit{Department of Information Engineering} \\
\textit{Department of Industrial Engineering} \\
\textit{University of Padua}\\
Padua, Italy \\
angelo.cenedese@unipd.it}
}

\maketitle

\begin{abstract}
The commissioning of industrial electric drives still relies heavily on manual tuning of cascaded control loops, requiring expert knowledge and significant time. In this paper, we propose a fully automated approach for tuning the current control loop of industrial drives using Bayesian Optimization (BO) directly on real hardware, without requiring a system model or firmware modifications. The drive is treated as a black-box system, and the controller parameters are iteratively updated through closed-loop experiments. The tuning problem is formulated as a multi-objective optimization task that directly minimizes tracking error, time-weighted error, overshoot, and oscillatory behavior, enabling the identification of Pareto-optimal controller configurations. To address discrete parameters, noisy evaluations, and limited budgets, we adopt a multivariate Tree-structured Parzen Estimator (TPE) as the underlying BO strategy. The proposed method operates under practical industrial constraints, including communication latency and limited evaluation budgets. 
The experimental validation on a real motor drive system under no-load conditions shows that the method achieves performance comparable to expert tuning within a few minutes and without human intervention. Results show that Gaussian Process (GP)-based BO can yield highly competitive final solutions, but TPE-based BO is better aligned with this setting due to faster convergence, richer Pareto-front approximation, and lower computational overhead.
\end{abstract}


\section{Introduction}

The commissioning of industrial electric drives remains a critical and 
time-consuming step in automation workflows. In particular, tuning the 
cascaded control loops --- including current, velocity, and position 
regulators --- requires careful adjustment of controller parameters to ensure stability, fast response, and low oscillations. In industrial practice, this process is often performed manually by expert technicians through heuristics and iterative trial-and-error, making it difficult to standardize and scale across different machines and operating conditions.
Data-driven and optimization-based methods have been proposed to automate controller tuning. Among them, Bayesian Optimization (BO) is particularly attractive for expensive black-box problems, since it aims to identify high-quality solutions with a limited number of evaluations. This makes it well-suited to control applications where each evaluation corresponds to a physical experiment. However, many existing BO-based approaches for controller tuning rely on simulations or simplified models, while direct validation on real industrial hardware under practical deployment constraints remains limited in the literature. 
In this setting, GP-based BO struggles because the parameter space is discrete, evaluations are noisy, and the budget is limited to a few tens of real-hardware trials. The Tree-structured Parzen Estimator (TPE) avoids these assumptions by modeling the parameter distribution non-parametrically, making it a more suitable choice for this setting.

In this paper, we propose a fully automated framework for the tuning of the current control loop of industrial drives directly on real hardware. The approach is model-free, requires no firmware modification, and operates through a standard communication interface. The tuning problem is cast as a multi-objective optimization task that directly minimizes four complementary performance metrics: tracking error, time-weighted error, overshoot, and oscillatory behavior. This formulation enables the identification of Pareto-optimal controller configurations without relying on scalarization during optimization.
Experimental validation on a real motor drive system shows that the proposed method can identify high-quality controller parameters within a limited number of trials, achieving competitive performance with expert tuning. The goal is not to claim uniform superiority on every metric, but to identify the BO strategy best aligned with industrial commissioning constraints. 

The main contributions of this paper are as follows:
\begin{itemize}
    \item A fully automated framework for tuning current control loops of industrial drives using BO directly on real hardware, without requiring system models or firmware modifications.
    \item A multi-objective formulation based on multivariate TPE that 
    directly optimizes multiple control performance metrics and is robust to discrete parameters and noisy hardware evaluations.
    \item An experimental validation demonstrating fast convergence, 
    competitive performance with expert tuning, robustness across repeated runs, and a comparison against Random Search (RS) and GP-based BO under the same evaluation budget.
\end{itemize}

\section{Related Work}

The tuning of proportional--integral/proportional--integral--derivative (PI/PID) controllers in industrial drives is still often performed through manual or heuristic procedures, including classical rules such as Ziegler--Nichols and Cohen--Coon, relay-based autotuning methods, and iterative trial-and-error by expert technicians~\cite{ziegler1942optimum, cohen1953theoretical, aastrom1984automatic}. While widely used thanks to their simplicity, these approaches are time-consuming, difficult to standardize, and strongly dependent on human expertise, often yielding suboptimal or oscillatory responses in complex systems.

A large body of literature has investigated PID tuning methods across different domains~\cite{borase2021review}. Existing reviews categorize these approaches into conventional rules, model-based techniques, and data-driven or optimization-based methods, each characterized by different trade-offs in performance, robustness, and computational complexity.

Among classical autotuning techniques, relay-feedback methods introduced by Åström and Hägglund~\cite{aastrom1984automatic} enable estimation of critical system characteristics directly from closed-loop experiments without requiring explicit models. These ideas have also inspired application-specific autotuning procedures, including current-loop autotuning for electric drives~\cite{pasqualotto2020model}. However, these approaches rely on predefined excitation procedures and analytical tuning rules rather than direct optimization of multiple performance criteria.

To improve performance and reduce manual effort, optimization-based methods have been widely explored. Evolutionary approaches such as Genetic Algorithms and Particle Swarm Optimization show promising results, particularly in simulation~\cite{gaing2004particle, HERREROS2002457}, but typically require a large number of evaluations, limiting their applicability when experiments are performed on real hardware. More generally, intelligent and nature-inspired techniques improve control performance but introduce additional complexity and computational cost~\cite{patil2024review}.

BO has recently emerged as a sample-efficient alternative for expensive black-box controller tuning. It has been successfully applied to digital multi-loop PID tuning~\cite{COUTINHO2023108211}, HVAC regulation~\cite{ijcai2019p811}, and safety-aware cascade control~\cite{khosravi2022safety}, with experimental validation on real motor systems~\cite{fujimoto2023controller}. Recent approaches also incorporate digital twins to further reduce hardware experiments~\cite{nobar2024guided}, while extensions to nonlinear and adaptive control have been explored in robotics~\cite{hajieghrary2022bayesian}.

Despite these advances, few works address the combination of constraints considered here: direct tuning on real industrial hardware, absence of plant models or digital twins, integer-valued parameters, noisy measurements, and strict limits on the number of evaluations. Moreover, most existing approaches focus on single-objective formulations or simplified experimental settings.

In contrast, this work formulates current-loop tuning in an industrial drive as a direct multi-objective black-box optimization problem executed entirely on real hardware. The proposed method does not rely on predefined tuning rules, system identification, or modeling assumptions, and enables efficient and fully automated commissioning directly on real hardware under realistic industrial constraints.

\section{Method}
\label{sec:method}

\subsection{Problem Formulation}

The goal of the proposed approach is to automatically tune the 
PI parameters of the current control loop of an industrial electric drive. The controller tuning problem is formulated as a black-box multi-objective problem, where each evaluation is performed on real hardware.

Let $\theta = (K_p, K_i)$ denote the controller parameters. For a given configuration, the system is excited using a predefined reference signal, and the corresponding response is acquired through the drive oscilloscope.

Instead of aggregating performance into a single scalar objective, we 
directly consider multiple performance metrics:
\begin{equation}
\mathbf{f}(\theta) =
\left[
\mathrm{IAE}(\theta),\;
\mathrm{ITAE}(\theta),\;
\mathrm{OS}(\theta),\;
\mathrm{OSC}(\theta)
\right]
\end{equation}

The objective is to simultaneously minimize all components:
\begin{equation}
\theta^* \in \arg\min_{\theta \in \mathcal{D}} \mathbf{f}(\theta)
\end{equation}
where $\mathcal{D}$ is the feasible parameter space. The solution is 
therefore a set of Pareto-optimal configurations representing trade-offs between tracking accuracy, transient response, overshoot, and oscillatory behavior. Accordingly, the optimization aims at approximating the Pareto set of non-dominated controller configurations within the feasible domain.

All controller parameters are constrained to integer values within 
predefined bounds, reflecting implementation requirements and the limited parameter resolution of the target industrial hardware. Each function evaluation corresponds to a physical experiment, making the optimization problem expensive and subject to measurement noise. This setting reflects practical industrial commissioning, where each trial incurs non-negligible time and hardware wear, making sample efficiency a key requirement. In the present study, all evaluations are conducted under no-load conditions to isolate the inner-loop electrical dynamics.

\subsection{BO with Multivariate TPE}

To efficiently explore the parameter space under a limited evaluation 
budget, we employ BO based on TPE~\cite{bergstra2011algorithms}, using its multivariate extension.

Classical GP-based BO provides a principled probabilistic surrogate but struggles in our setting due to three concurrent factors: (i)~the parameter space is discrete, (ii)~hardware evaluations are inherently noisy, and (iii)~the budget is limited to a few tens of trials with four simultaneous objectives. TPE addresses these limitations by modeling the parameter distribution non-parametrically, without assuming smoothness or continuity of the objective landscape~\cite{bergstra2011algorithms,watanabe2023tree}.

In the standard TPE formulation, each parameter is modeled independently. In contrast, the multivariate variant jointly models the parameter distribution, capturing dependencies between $K_p$ and $K_i$~\cite{optuna_2019}. This is particularly important in control applications, where parameters are strongly coupled and jointly determine system stability and performance.

For the multi-objective setting, TPE maintains a set of non-dominated 
solutions and guides the search toward regions of the parameter space that improve the Pareto front. At each iteration, candidate parameters are sampled by favoring configurations that are more likely to dominate previously observed solutions.

Multivariate TPE is well suited to this application because it handles noisy discrete evaluations while capturing parameter interactions without requiring a continuous surrogate. In the experiments, the first $n_0 = 10$ startup trials are sampled randomly before switching to the model-based proposal strategy, so that $N_{\mathrm{total}} = n_0 + N_{\mathrm{BO}}$ where $N_{\mathrm{BO}}$ denotes the number of model-guided trials.

\subsection{Experimental Evaluation Loop}

The optimization is executed through an external loop interacting with the drive via a communication interface. Controller parameters are updated remotely by writing to specific memory addresses of the drive, whose locations are retrieved from a hardware configuration file. This mechanism allows seamless interaction with the drive firmware without requiring any modification to it. Each trial consists of the following steps:

\begin{enumerate}
    \item \textbf{Parameter update:} the candidate $(K_p, K_i)$ values are written to the corresponding controller registers.
    \item \textbf{Excitation:} a predefined reference signal is applied.
    \item \textbf{Data acquisition:} reference and response signals are collected using the drive oscilloscope.
    \item \textbf{Metric computation:} the performance metrics are computed from the acquired signals.
\end{enumerate}

All trials are conducted under identical operating conditions to ensure consistency and comparability.

\subsection{Performance Metrics}

Each controller configuration is evaluated through four complementary
metrics computed from the reference signal $r(t)$ and the measured
response $y(t)$.

\paragraph{Preprocessing}
All metrics are normalized by $d = \max(r) - \min(r)$ to ensure 
comparability across excitation profiles of different amplitudes. In the considered experiments, the excitation always contains a nonzero step, hence $d > 0$.
The overshoot and oscillation metrics are evaluated on the 
individual constant-reference segments of the excitation profile.

\paragraph{Global tracking metrics}
Let $e_i = r_i - y_i$ denote the pointwise tracking error. Global tracking quality is quantified by two complementary objectives:

\begin{equation}
\mathrm{IAE} = \frac{1}{N\,d} \sum_{i=0}^{N-1} |e_i|
\end{equation}
\begin{equation}
\mathrm{ITAE} = \frac{1}{d} \sum_{i=0}^{N-1} \frac{i}{N-1}\,|e_i|
\end{equation}

IAE captures average tracking accuracy, while ITAE weights errors 
by their time index, penalizing deviations that persist later in 
the response and thus discriminating between fast and slow settling behavior.

\paragraph{Directional overshoot}
For each transition $k$, let $\delta_k = r_{k+1} - r_k$ be the 
amplitude, $\sigma_k = \mathrm{sign}(\delta_k)$, and $[s_{k+1}, f_{k+1}]$ the index range of the constant-reference segment immediately following the transition. The local overshoot is:

\begin{equation}
\mathrm{OS}_k = \max\!\left(0,\;
    \frac{
        \displaystyle\max_{i \in [s_{k+1},\, f_{k+1}]}
        \bigl(\sigma_k\cdot y_i\bigr)
        \;-\; \sigma_k \cdot r_{k+1}
    }{|\delta_k|}
\right)
\end{equation}

Projecting through $\sigma_k$ handles both positive and negative steps consistently. The global overshoot is the worst case:

\begin{equation}
    \mathrm{OS} = \max_k\; \mathrm{OS}_k
\end{equation}

\paragraph{Oscillation penalty}
Residual oscillatory behavior is quantified as the normalized 
RMS of the zero-mean tracking error on each constant-reference 
segment of length $L$:

\begin{equation}
\mathrm{OSC}_{\mathrm{seg}} =
    \frac{1}{d}\sqrt{\frac{1}{L}\sum_{i=0}^{L-1}(e_i - \bar{e})^2}
\end{equation}

where $\bar{e}$ is the segment mean, subtracted to isolate genuine 
ripple from steady-state offset. Segments are classified as 
\emph{active} or \emph{zero} depending on the reference level;
$\mathrm{OSC}_{\mathrm{active}}$ and $\mathrm{OSC}_{\mathrm{zero}}$ 
denote the worst case over each class, and the final metric is:

\begin{equation}
    \mathrm{OSC} = \max(\mathrm{OSC}_{\mathrm{active}},\;
                        \mathrm{OSC}_{\mathrm{zero}})
\end{equation}

This makes the metric sensitive to residual ripple both during 
active tracking and around zero-current conditions, which are 
both operationally relevant in industrial drives.

\subsection{Selection of Final Controller}

Since the optimization produces a set of Pareto-optimal solutions,
a single controller configuration must be selected for deployment.
Selection is performed in two sequential steps.

\paragraph{Step 1 --- Hard constraints}
If upper bounds are imposed on overshoot and/or oscillation, only the Pareto-front trials satisfying all active constraints are retained as candidates. If no trial satisfies the constraints, the full Pareto front is used as a fallback.

\paragraph{Step 2 --- Weighted distance from the ideal point}
Let $\mathcal{C}$ denote the candidate set after Step~1, and let $\mathbf{v}_j = (\mathrm{IAE}_j, \mathrm{ITAE}_j, \mathrm{OS}_j, \mathrm{OSC}_j)$ be the objective vector of candidate $j \in \mathcal{C}$. Each metric is normalized to $[0,1]$ across $\mathcal{C}$:
\begin{equation}
    \tilde{v}_{j,m} = \frac{v_{j,m} - \min_k v_{k,m}}
                          {\max_k v_{k,m} - \min_k v_{k,m}}
\end{equation}
where the denominator is set to $1$ if all candidates share the same value for metric $m$. The selected controller is the candidate with the smallest weighted $\ell_2$ distance from the ideal point $\tilde{\mathbf{v}} = \mathbf{0}$, which corresponds to the best achievable value for each normalized metric:

\begin{equation}
    j^* = \arg\min_{j \in \mathcal{C}}\;
    \sqrt{\sum_{m} w_m \,\tilde{v}_{j,m}^2}
\end{equation}

The weight vector $\mathbf{w} = (w_{\mathrm{IAE}},\, 
w_{\mathrm{ITAE}},\, w_{\mathrm{OS}},\, w_{\mathrm{OSC}})$ encodes the deployment priority and can be adjusted to reflect application-specific requirements without re-running the optimization. Three predefined strategies are provided:
\begin{itemize}
    \item \textbf{Balanced}: equal importance to all objectives;
    \item \textbf{Fast}: prioritizes tracking speed (IAE, ITAE), tolerating moderate overshoot and oscillation;
    \item \textbf{Smooth}: prioritizes transient quality (overshoot, oscillation), tolerating slower convergence.
\end{itemize}

This post-selection mechanism decouples optimization from deployment priorities. In the present single-loop setting, differences across strategies are modest, but the same mechanism can support future extensions to outer loops.

\section{Experimental Setup}

\subsection{Hardware Setup and Control Architecture}

\begin{figure}[t]
    \centering
    \includegraphics[width=0.8\linewidth]{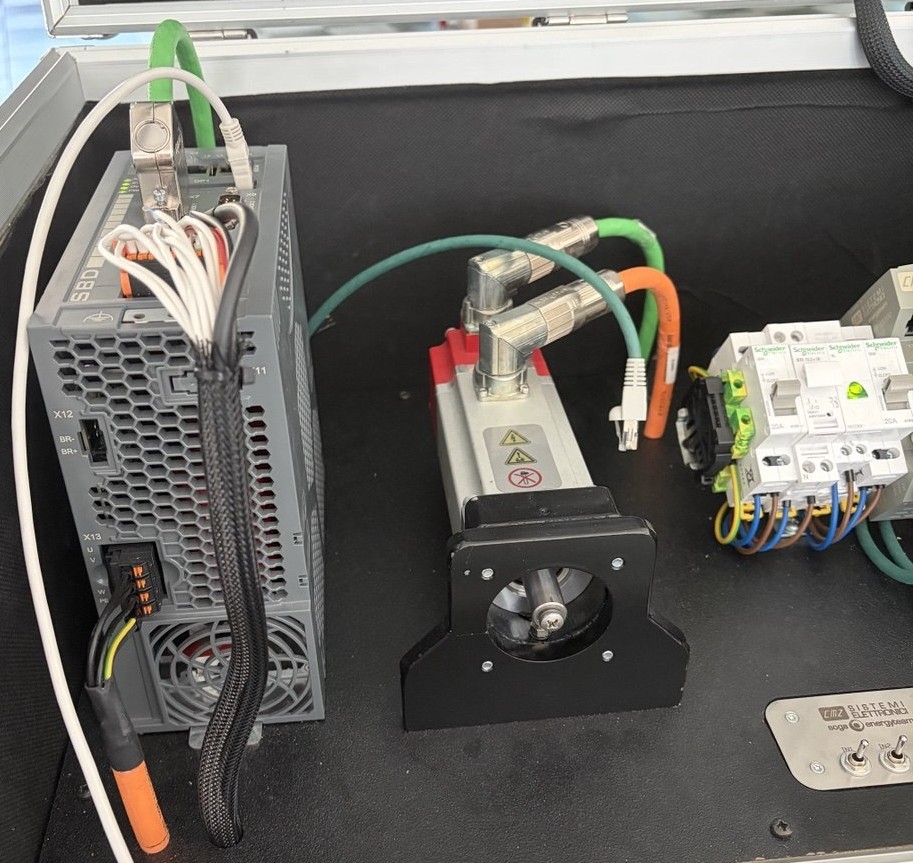}
    \caption{Experimental setup used for current-loop tuning on the industrial drive system.}
    \label{fig:hardware_setup}
\end{figure}

The proposed method is validated on a real industrial electric drive 
system, as shown in Fig.~\ref{fig:hardware_setup}. The setup consists 
of an industrial motor drive connected to a brushless AC motor 
(230\,V, 6-pole, 1.8\,A rated current, 3000\,rpm rated speed) and 
controlled through an external PC.

The optimization runs on the external PC and communicates with the 
drive via Modbus. Signal acquisition relies on a proprietary 
high-speed variant (Fast Modbus) that reduces communication overhead; 
the framework remains compatible with standard Modbus RTU at the cost 
of a moderately longer per-trial acquisition time. Controller 
parameters are updated by writing to specific memory addresses 
retrieved from a hardware configuration file, making the system 
firmware-independent and portable across drive configurations without 
requiring any firmware modification.

The drive adopts a standard cascaded control architecture. This work focuses exclusively on the current control loop---the innermost and fastest loop in the hierarchy---tuning the proportional and integral gains $(K_p, K_i)$ of the current loop PI controller, while outer loops (velocity, position) are left unchanged. All experiments are conducted under no-load conditions, which isolates the electrical dynamics of the current loop and provides a controlled, repeatable benchmark. Specifically, FOC maps the stator currents into a rotating reference frame aligned with the rotor flux, yielding two decoupled components: the $d$-axis current regulates the flux level and does not contribute to torque production, while the $q$-axis current is associated with electromagnetic torque generation. The tuning targets the $d$-axis current loop, while the $q$-axis reference is held at zero throughout all experiments. This choice ensures that no torque is generated during the tuning procedure, providing safe and repeatable excitation conditions while preserving the same electrical dynamics as the $q$-axis loop.

\subsection{Data Acquisition and Excitation Signal}

The system is excited using a rectangular pulse reference profile consisting of a positive current step followed by a return to zero, enabling simultaneous evaluation of overshoot, tracking accuracy, and steady-state oscillations under a single excitation. The excitation amplitude is set to 60\% of the rated peak current, i.e., $I_{\mathrm{rated}} \times \sqrt{2}  \times 0.6 \approx 1.53\,\mathrm{A}$, and the pulse duration is 40\,ms. The amplitude and duration are kept constant across all trials to ensure comparability.

Responses are acquired via the internal oscilloscope of the drive, 
which synchronously records both the reference and the measured current signal at a fixed sampling time of $50\,\mu$s. All signals are 
processed offline after each trial to compute the performance metrics 
defined in Section~\ref{sec:method}, and all experimental data are 
logged for reproducibility.

\subsection{Optimization Setup}

The optimization process is performed over the controller parameters 
$(K_p, K_i)$, which are constrained within predefined bounds, i.e., 
$K_p, K_i \in [500, 10000]$, based on practical considerations derived 
from the drive specifications. Both parameters are treated as 
integer-valued variables, reflecting the limited parameter resolution of the industrial hardware.

Optimization is evaluated at 15, 20, 30, 50, and 100 trials, with 5 independent seeds per budget.

The total tuning time scales approximately linearly with the number of 
trials, ranging from about 1.5~minutes for 15~trials to 12~minutes for 
100~trials for both TPE-based BO and RS. This is because per-trial cost is dominated by hardware communication and signal acquisition, making the computational overhead of the optimization algorithm negligible.

In contrast, GP-based BO introduces a non-negligible computational overhead due to surrogate model fitting and acquisition optimization. As a result, the total tuning time increases up to approximately 18~minutes for 100~trials.

All methods are compared under identical operating conditions and equal real-hardware evaluation budgets (same number of trials).

\subsection{Baseline and Evaluation Metrics}

Three references are considered: expert tuning, RS, and GP-based BO. First, a set of expert-tuned parameters provided by experienced technicians serves as a practical performance reference. Second, a RS baseline is used to isolate the contribution of the TPE-based acquisition strategy from the benefit of simply sampling the parameter space under the same evaluation budget. Third, GP-based BO is included as a representative model-based optimization approach to assess the impact of different surrogate modeling strategies under the same experimental conditions.

The optimization methods are compared in terms of four performance metrics (IAE, ITAE, OS, OSC) computed on the selected Pareto-front solution, as well as the number of Pareto-optimal solutions identified. To assess convergence quality across the full optimization trajectory, the hypervolume indicator~\cite{10.1007/BFb0056872} is also tracked over trials. To ensure a fair comparison, all objective values are normalized to $[0,1]$ using bounds computed globally across all runs and methods, and a fixed reference point is set to the worst observed values after global normalization across all methods and runs.

\subsection{Implementation Details}

The optimization loop is implemented in Python on the external computing unit, using the Optuna framework~\cite{optuna_2019}. The overall pipeline, including parameter updates and signal acquisition, is fully automated and implemented in Python. Communication with the drive is handled through a custom Python module developed for this work, which interfaces with the drive via Modbus and retrieves hardware-specific memory addresses from a configuration file. This setup enables a realistic evaluation of the proposed approach in a practical industrial scenario and is directly deployable on standard industrial hardware without any firmware modification.

\section{Experimental Results}

\subsection{Evaluation Protocol}

The proposed method is evaluated across multiple trial budgets (15, 20, 30, 50, and 100 trials), with 5 independent runs per configuration to assess robustness and repeatability.

All results reported in this section are obtained using the \textbf{Balanced} selection strategy (equal weights on all four objectives), without Step~1 filtering. The analysis focuses on the convergence behavior of the optimization process, the structure of the obtained Pareto fronts, and the comparison of different optimization methods under identical evaluation budgets.

\subsection{Optimization Performance and Pareto-Front Quality}

\paragraph{Convergence analysis}
The convergence behavior of the optimization process is evaluated by tracking the evolution of the normalized hypervolume over the optimization trials. The hypervolume measures the volume of the objective space dominated by the current Pareto front with respect to a fixed reference point, thus capturing both the quality and the diversity of the identified solutions. Higher values indicate a better approximation of the Pareto front.

All methods share a common initialization phase: the first $n_0 = 10$ trials are sampled randomly to collect an initial dataset before model-based optimization is activated. As a result, similar performance is expected during this phase.

\begin{figure}[t]
    \centering
    \includegraphics[width=\linewidth]{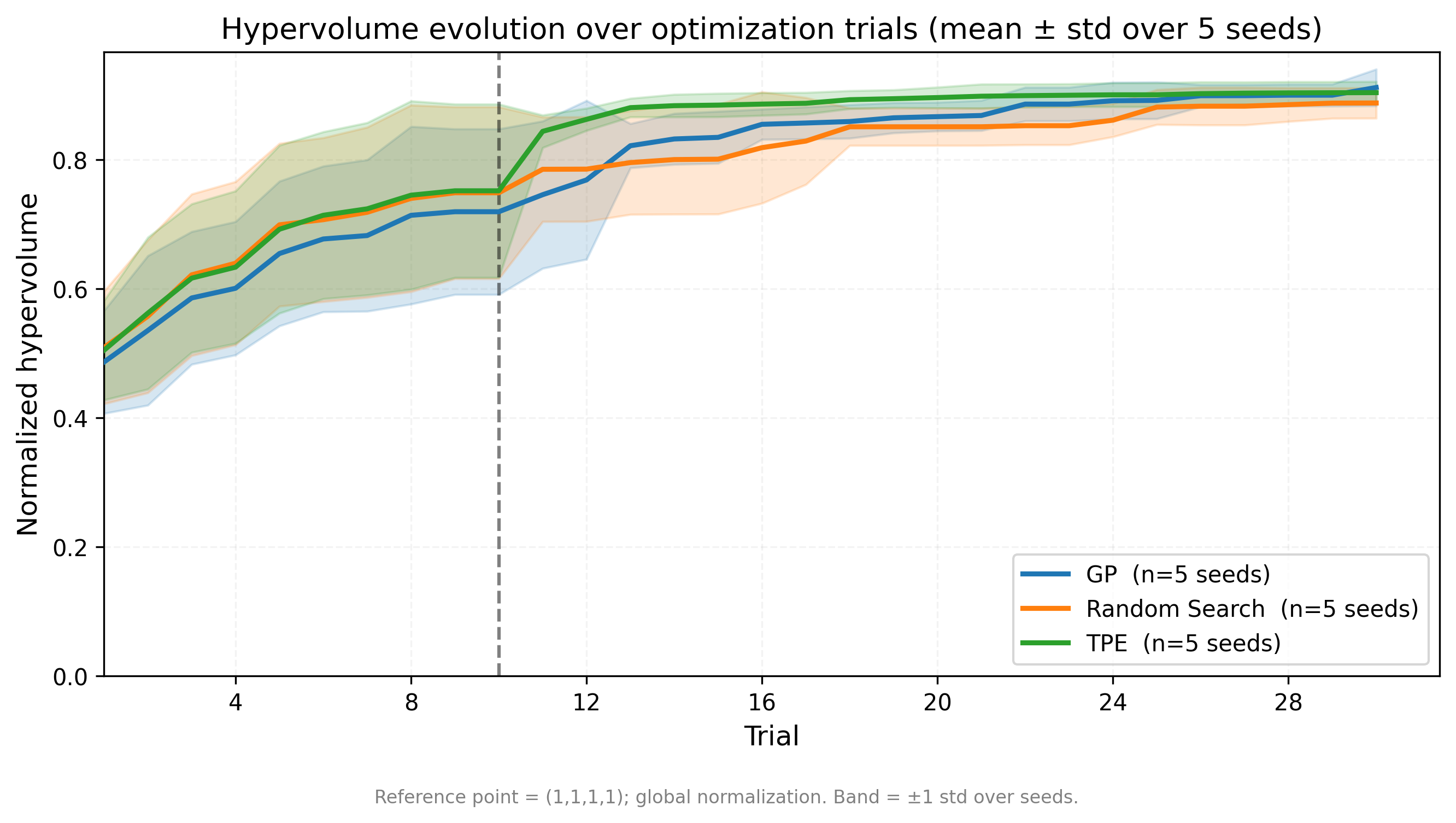}
    \caption{Normalized hypervolume over trials for TPE-based BO, GP-based BO, and RS (mean $\pm$ std over 5 seeds). The dashed line marks the end of the random initialization phase.}
    \label{fig:hypervolume}
\end{figure}

Fig.~\ref{fig:hypervolume} shows the evolution of the normalized hypervolume for TPE-based BO, GP-based BO, and RS, averaged over 5 independent seeds with the corresponding standard deviation band. Consistent with the shared initialization phase, all methods exhibit similar performance during the first 10 trials. Clear differences emerge once model-based sampling is activated. In particular, TPE exhibits faster hypervolume growth compared to both GP-based BO and RS, especially immediately after the transition from random exploration to model-guided optimization. This is particularly relevant under limited evaluation budgets. In practical commissioning, this translates into faster tuning procedures and reduced system downtime.

GP-based BO provides a competitive alternative, often improving over RS and in some cases yielding very strong final selected solutions, although it does not match the early hypervolume growth of TPE. This can be attributed to the difficulty of modeling discrete parameter spaces and noisy hardware evaluations with GP-based BO in a low-budget regime. In contrast, the non-parametric formulation of TPE is better suited to this setting, enabling more effective exploration once sufficient observations are available.

At later stages, all methods approach similar hypervolume values, but TPE retains an advantage in convergence speed and variability.

While the final selected solutions achieve comparable performance across methods (Table~\ref{tab:bo_random_mean_std}), clear differences emerge in search efficiency and Pareto front structure. In particular, both GP-based BO and TPE-based BO identify larger Pareto fronts than RS as the evaluation budget increases, confirming the benefit of model-based guidance.

\begin{table*}[t]
\centering
\caption{Results across trial budgets for TPE-based BO, GP-based BO, and RS (mean $\pm$ std over 5 seeds). Selected solutions use the Balanced strategy; bold indicates the lowest mean value per budget and metric; expert tuning is shown for reference.}
\label{tab:bo_random_mean_std}
\small
\setlength{\tabcolsep}{6pt}
\begin{tabular}{llccccc}
\toprule
Method & Trials & IAE & ITAE & OS & OSC & Pareto size \\
\midrule
RS  & 15  & $0.00779\pm0.00135$ & $8.52\pm0.99$  & $0.0815\pm0.0646$ & $0.0674\pm0.0056$ & $5.4\pm1.1$  \\
GP-based BO  & 15  & $0.00703\pm0.00030$ & $8.11\pm0.24$ & $0.0429\pm0.0261$ & $0.0670\pm0.0024$ & $5.6\pm1.5$ \\
TPE-based BO & 15  & $\textbf{0.00686}\pm0.00023$ & $\textbf{7.82}\pm0.22$  & $\textbf{0.0318}\pm0.0183$ & $\textbf{0.0662}\pm0.0009$ & $\textbf{5.8}\pm2.2$  \\
\midrule
RS  & 20  & $0.00730\pm0.00113$ & $8.27\pm0.88$  & $0.0886\pm0.0611$ & $\textbf{0.0645}\pm0.0024$ & $6.6\pm2.5$  \\
GP-based BO  & 20  & $0.00686\pm0.00023$ & $8.00\pm0.17$ & $0.0496\pm0.0228$ & $0.0686\pm0.0059$ & $6.6\pm1.1$ \\
TPE-based BO & 20  & $\textbf{0.00685}\pm0.00016$ & $\textbf{7.93}\pm0.17$  & $\textbf{0.0394}\pm0.0337$ & $0.0651\pm0.0023$ & $\textbf{8.2}\pm1.8$  \\
\midrule
RS  & 30  & $\textbf{0.00676}\pm0.00019$ & $7.98\pm0.28$  & $0.0713\pm0.0222$ & $\textbf{0.0614}\pm0.0047$ & $7.4\pm3.0$  \\
GP-based BO  & 30  & $0.00690\pm0.00018$ & $8.10\pm0.15$ & $\textbf{0.0430}\pm0.0104$ & $0.0627\pm0.0047$ & $11.2\pm2.2$ \\
TPE-based BO & 30  & $\textbf{0.00676}\pm0.00027$ & $\textbf{7.76}\pm0.18$  & $0.0451\pm0.0201$ & $0.0646\pm0.0008$ & $\textbf{11.6}\pm0.5$ \\
\midrule
RS  & 50  & $0.00688\pm0.00030$ & $\textbf{7.92}\pm0.30$  & $0.0529\pm0.0265$ & $0.0652\pm0.0019$ & $9.2\pm2.5$  \\
GP-based BO  & 50  & $\textbf{0.00677}\pm0.00022$ & $7.98\pm0.18$ & $\textbf{0.0418}\pm0.0210$ & $0.0627\pm0.0050$ & $\textbf{20.4}\pm4.4$ \\
TPE-based BO & 50  & $0.00714\pm0.00026$ & $8.16\pm0.19$  & 
$0.0499\pm0.0364$ & $\textbf{0.0623}\pm0.0034$ & $\textbf{20.4}\pm1.5$ \\
\midrule
RS  & 100 & $0.00683\pm0.00037$ & $7.94\pm0.34$  & $0.0440\pm0.0218$ & $0.0647\pm0.0019$ & $16.2\pm3.8$ \\
GP-based BO  & 100 & $\textbf{0.00669}\pm0.00012$ & $\textbf{7.84}\pm0.12$ & $\textbf{0.0293}\pm0.0141$ & $0.0635\pm0.0040$ & $29.2\pm9.7$ \\
TPE-based BO & 100 & $0.00728\pm0.00044$ & $8.26\pm0.40$  & $0.0347\pm0.0191$ & $\textbf{0.0606}\pm0.0042$ & $\textbf{38.2}\pm9.3$ \\
\midrule
Expert & -- & 0.00967 & 10.62 & 0.02143 & 0.08123 & -- \\
\bottomrule
\end{tabular}
\end{table*}

\begin{figure*}[t]
    \centering
    \includegraphics[width=\textwidth]{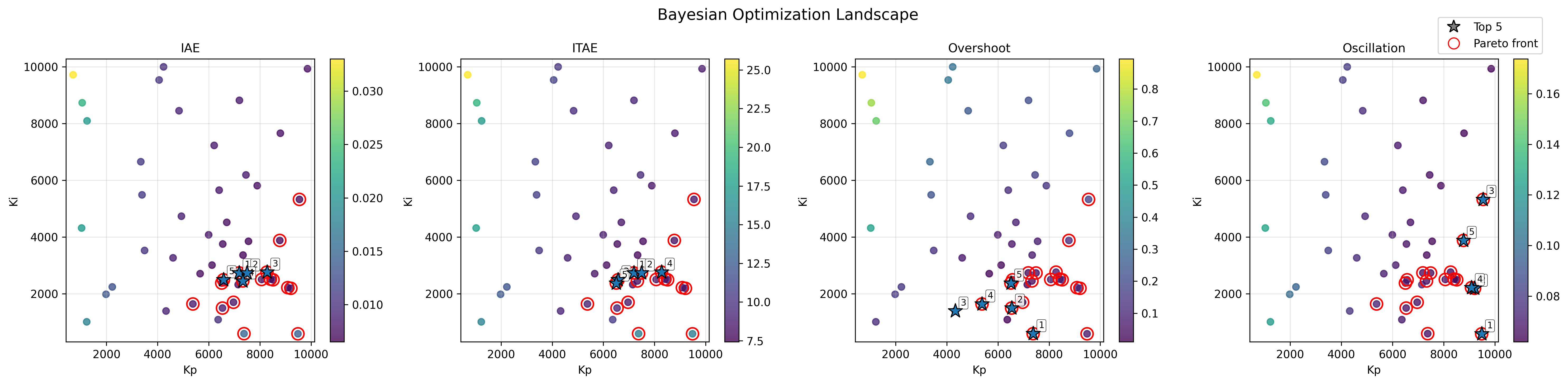}
    \caption{Evaluated configurations in the $(K_p, K_i)$ space, color-coded by objective value. Pareto-optimal solutions and the top-5 configurations per metric are highlighted. Representative run after 30 trials (TPE-based BO).}
    \label{fig:pareto}
\end{figure*}

\paragraph{Pareto-front structure and trade-offs}
The multi-objective formulation enables the identification of a set of Pareto-optimal solutions that capture the trade-offs among the considered performance metrics. Fig.~\ref{fig:pareto} shows all configurations evaluated in a representative optimization run, plotted in the parameter space $(K_p, K_i)$ and color-coded by objective value, with Pareto-optimal solutions and the top-5 configurations per metric highlighted.

The results highlight clear trade-offs between tracking performance and transient behavior. Configurations achieving low tracking error (IAE, ITAE) may exhibit higher overshoot or oscillatory behavior, while more conservative parameter settings reduce transient excursions at the cost of slower convergence. This confirms that no single configuration simultaneously minimizes all objectives, and validates the use of a multi-objective formulation.

Table~\ref{tab:bo_random_mean_std} also reports the average final Pareto-front size, which quantifies how effectively each method explores the trade-off surface. At low trial budgets, all methods yield comparable Pareto front sizes, consistent with the shared initialization phase. As the number of trials increases, model-based methods significantly outperform RS. In particular, both GP-based BO and TPE-based BO identify larger Pareto sets, suggesting a richer empirical approximation of the trade-off surface, especially when considered jointly with the hypervolume trends.

Among the considered approaches, TPE consistently achieves the largest empirical Pareto set across all budgets. This effect becomes particularly evident at higher trial counts, where TPE identifies substantially more non-dominated solutions than both GP-based BO and RS. This provides a richer empirical trade-off representation for controller selection.

Overall, these results indicate that the primary advantage of model-based optimization lies not only in improving the best solution, but also in efficiently approximating the Pareto front. While GP-based BO can yield highly competitive final selected controllers, TPE-based BO emerges as the more suitable overall choice for this industrial multi-objective setting, due to its better early-stage convergence, richer empirical Pareto-front reconstruction, and lower computational overhead.

\subsection{Robustness Across Seeds}

Table~\ref{tab:bo_random_mean_std} shows that RS exhibits the largest variability across seeds, particularly at low trial budgets. This is especially evident for IAE and, even more markedly, for overshoot, whose standard deviation remains high at 15 and 20 trials. Such behavior reflects the unguided nature of random exploration, which makes the final selected solution strongly dependent on the specific sampled configurations.

Both BO-based methods provide more stable behavior across seeds than RS. GP-based BO often yields the lowest standard deviation on the selected-solution metrics, especially for IAE, ITAE, and OS at medium and high trial budgets, indicating strong consistency once sufficient data are available to fit the surrogate model.

TPE-based BO, while not always achieving the lowest variability on every individual metric, shows competitive stability across seeds and remains consistently more robust than RS. In particular, TPE exhibits very low variability on OSC at several budgets and, more importantly, a consistent Pareto-front size across repeated runs at intermediate budgets, e.g., $11.6\pm0.5$ at 30 trials and $20.4\pm1.5$ at 50 trials. This is a relevant property in the present multi-objective setting, where robustness should be evaluated not only in terms of the final selected controller, but also in terms of the consistency of the overall Pareto-front approximation across repeated runs.

These observations are consistent with the hypervolume analysis in Fig.~\ref{fig:hypervolume}, where BO-based methods exhibit more regular convergence behavior than RS. In particular, TPE combines strong convergence performance with a consistently richer set of non-dominated solutions, which is especially important in practical deployment scenarios where the goal is not only to obtain one good controller, but also to reliably recover a diverse set of high-quality trade-offs.

Overall, the results suggest that BO substantially improves repeatability over RS, while TPE-based BO offers the most compelling overall balance between convergence efficiency, robustness, and Pareto-front consistency under limited evaluation budgets.

A closer inspection of Table~\ref{tab:bo_random_mean_std} reveals that the IAE and ITAE values selected by TPE-based BO increase slightly at 50 and 100 trials compared to 30 trials. This behavior is consistent with the growth of the Pareto front: as more non-dominated solutions are identified, the Balanced selection strategy increasingly favors configurations that achieve lower overshoot and oscillation at the cost of marginally higher tracking error. This trade-off is an expected consequence of applying uniform weighting across all four objectives on a richer and more diverse Pareto set, and should therefore be interpreted as an effect of the post-selection criterion rather than as evidence of a weaker search process.

\subsection{Comparison with Expert Tuning}

To assess the practical effectiveness of the proposed method, the solution selected from the final Pareto set is compared with a manually tuned baseline provided by experienced technicians. The expert configuration represents the quality achievable through standard commissioning practice and serves as the primary practical reference.

\begin{figure*}[t]
    \centering
    
    \subfloat[Tuning signal]{\includegraphics[width=0.49\linewidth]{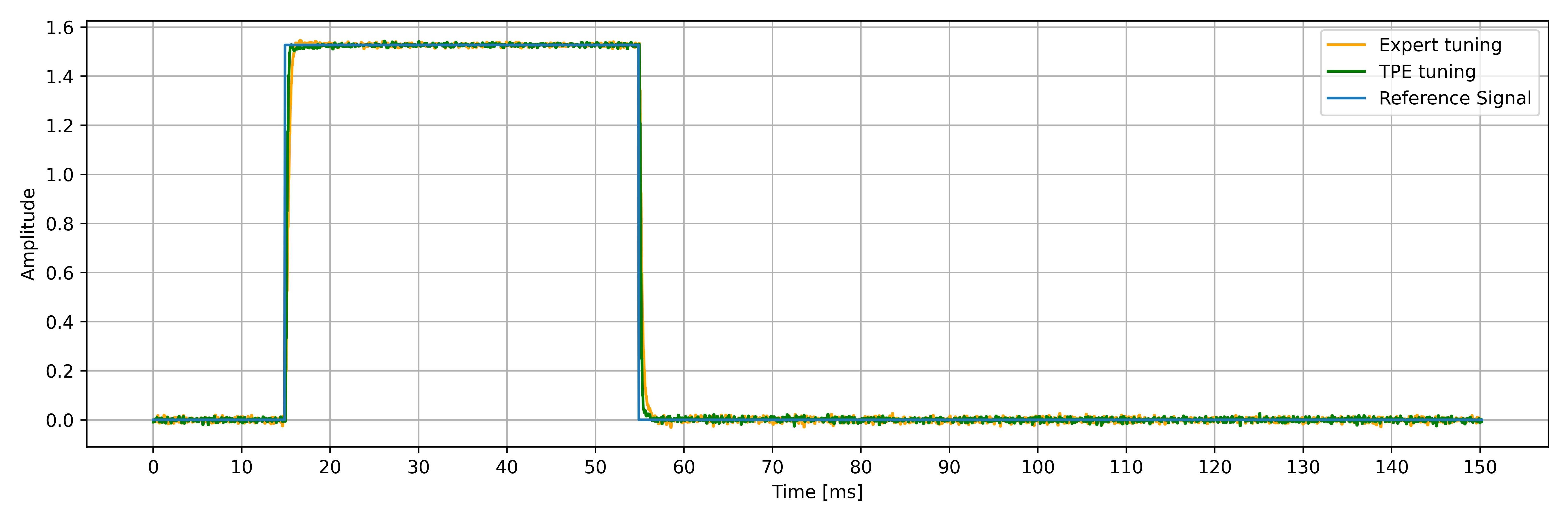}\label{fig:tuning_sub}}
        \hfill
    \subfloat[Validation signal]{\includegraphics[width=0.49\linewidth]{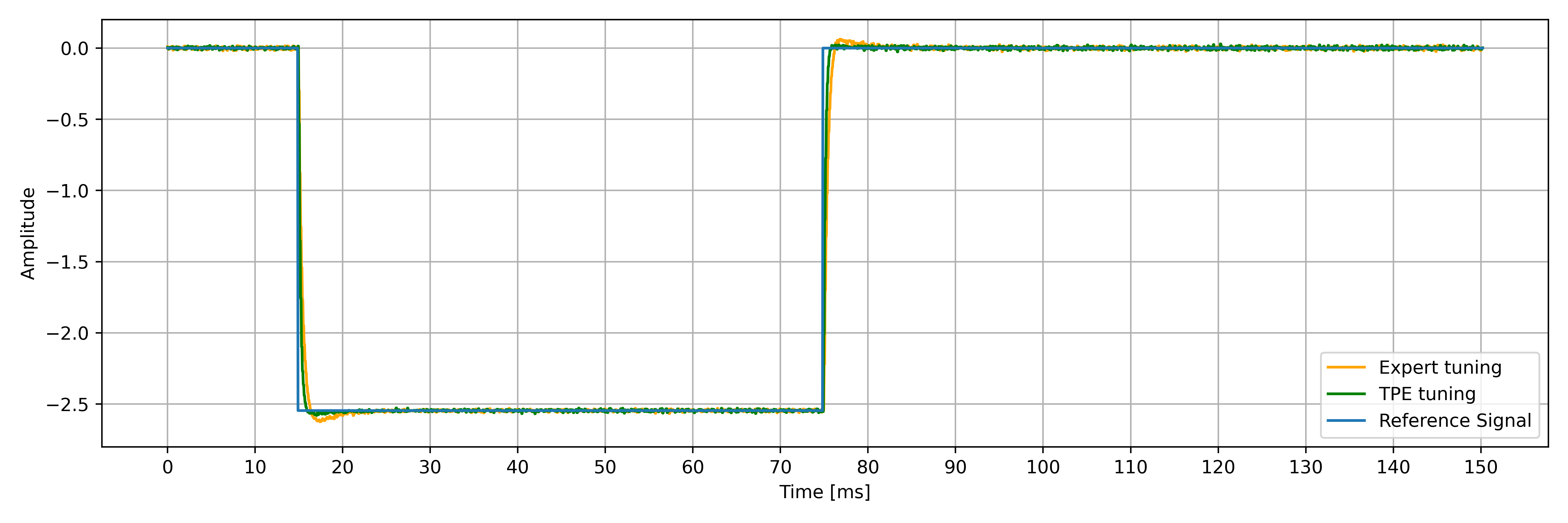}\label{fig:validation_sub}}

    \caption{Current-loop responses for the controller selected by TPE-based BO using the Balanced strategy and for the expert baseline. Left: tuning excitation. Right: validation excitation. Representative run after 30 trials.}
    \label{fig:tuning_combined}
\end{figure*}

Fig.~\ref{fig:tuning_sub} compares the current-loop response obtained with the expert-tuned parameters and with the controller selected by the proposed method. The optimized controller achieves performance comparable to expert tuning, and improves upon the expert baseline on several metrics: at 30 trials, TPE-based BO improves IAE by approximately 30\% and ITAE by 27\% relative to the expert baseline, while also reducing oscillatory behavior, as shown in Table~\ref{tab:bo_random_mean_std}. The expert configuration, on the other hand, achieves lower overshoot, reflecting a different trade-off among tracking accuracy, transient response, and oscillatory behavior.

Importantly, the automated procedure identifies competitive controller settings without any prior knowledge of the system, operating purely from closed-loop measurements. This result suggests that the proposed framework is capable of approaching and in several respects exceeding the quality of manual commissioning autonomously within a practical time budget.

\section{Conclusion}

The experimental results suggest that the proposed approach is effective for real-world tuning of industrial current control loops. Among the evaluated optimizers, TPE-based BO should not be interpreted as uniformly dominant over GP-based BO on every final metric; rather, it appears as the most practically suitable choice in this setting, thanks to its favorable balance between convergence speed, Pareto-front reconstruction quality, robustness, and computational overhead.

The method operates under realistic industrial constraints, including 
limited evaluation budgets, integer-valued parameters, noisy hardware 
measurements, and absence of system models, making it suitable for 
practical deployment. The optimization loop runs externally and requires no firmware modification, enabling seamless integration with existing industrial drive architectures. 

The availability of multiple Pareto-optimal configurations also supports adaptive or application-specific controller selection without re-running the optimization.

As shown in Fig.~\ref{fig:validation_sub}, the selected controller was additionally evaluated on a different validation excitation profile not used during tuning: a negative current step of amplitude $I_{\mathrm{rated}} \times \sqrt{2} \approx 2.55\,\mathrm{A}$ (i.e., $-$100\% of the rated peak current) with a pulse duration of 60\,ms. The optimized parameters preserved stable tracking and controlled transient response across both excitation profiles, providing preliminary evidence that the identified configurations are not overly tied to the specific tuning excitation signal.

A key limitation of the present study is that the experimental validation is restricted to the no-load current loop of a single motor-drive setup. Although suitable to isolate inner-loop electrical dynamics, broader validation under load and across multiple drive configurations is still needed.

Future work will extend the evaluation to loaded conditions, different drive configurations and outer control loops, where the interaction with mechanical dynamics becomes more significant. The proposed framework provides a reusable optimization and logging pipeline that naturally extends to these settings, and the collected data could support informed initialization, knowledge transfer across similar drive systems, or integration with digital twin environments for partial offline pre-tuning of controller parameters. 

These results represent a concrete step toward fully autonomous commissioning of industrial drive systems.

\section*{Acknowledgment}

The authors gratefully acknowledge CMZ Sistemi \mbox{Elettronici} S.r.l. for providing the industrial hardware platform that enabled the experimental validation of the proposed method, and thank their software engineering team for the support in the development of the communication interface and for the assistance during the experimental activities.

\bibliographystyle{IEEEtran}
\bibliography{references}

\end{document}